\documentclass[11pt, a4paper]{amsart}
\usepackage{amsmath}
\usepackage{amssymb}

\def\eq#1{(\ref{#1})}
\newtheorem{thm}{Theorem}
\newtheorem{prop}[thm]{Proposition}
\newtheorem{lemma}[thm]{Lemma}

\renewcommand{\leq}{\leqslant}
\renewcommand{\geq}{\geqslant}
\newcommand{\pf}{\noindent{\it Proof \ }}
\newcommand{\epf}{{$\quad$ \hfill $\Box$}}

\def\de{\delta}
\def\pa{\partial}
\def\Z{\mathbb{Z}}
\def\C{\mathbb{C}}

\def\cA{\mathcal{A}}

\def\cD{\mathcal{D}}
\def\cG{\mathcal{G}}

\def\bt{\bar{t}}
\def\bX{\bar{X}}
\def\bL{\bar{L}}
\def\bu{\bar{u}}

\def\L{\Lambda}

\def\im{\mathrm{Im}}
\def\Ker{\mathrm{Ker}}
\DeclareMathOperator{\res}{Res}
\DeclareMathOperator{\tr}{Tr}
\DeclareMathOperator{\diag}{diag}
\DeclareMathOperator{\End}{End}

\def\cF{\mathcal{F}}
\def\tPi{\tilde{\Pi}}
\def\bh{\bar{h}}

\begin{document}

\title[Two-dimensional Toda lattice]{The Hamiltonian structures of the two--dimensional Toda lattice and $R$-matrices}
\author{Guido Carlet}
\address{Department of Pure Mathematics and Mathematical Statistics, University of Cambridge, Cambridge CB3 0WB, United Kingdom.}
\email{g.carlet@dpmms.cam.ac.uk}
\begin{abstract}
We construct the tri-Hamiltonian structure of the two-dimensional Toda hierarchy using the $R$-matrix theory. 
\end{abstract}
\maketitle

\section{Introduction}

The two-dimensional Toda lattice is a well known system of differential-difference equations.
It was introduced by Mikhailov \cite{mikhailov79} in the form of equation \eq{i1}. The full hierarchy of commuting flows was constructed by Ueno and Takasaki in \cite{ueno-takasaki84}. Recently the interest in the two-dimensional Toda hierarchy has been stimulated by the discovery of its role in the problems of interface growth \cite{mineev2000}.

In this article we apply the classical $R$-matrix formalism to obtain the tri-Hamiltonian formulation of this hierarchy.  
The classical $R$-matrix method first appeared in the work on the quantum inverse-scattering method by Sklyanin \cite{sklyanin80}. 
It was further investigated from the integrable systems point of view by Semenov-Tian-Shanksy in \cite{semenov84}. 
Since that it becomes an important part of the Hamiltonian theory of integrable systems.

The Semenov-Tian-Shansky formulation was later adapted to the ``non-unitary'' case where the $R$-matrix is not skew-symmetric by Oevel-Ragnisco \cite{oevel-ragnisco89} and Li-Parmentier \cite{li-parmentier89}. 
The non-unitary case includes as main example various Toda-type hierarchies, considered in \cite{oevel-ragnisco89} and in subsequent works \cite{oevel97}.
The Hamiltonian structures for such hierarchies were also studied by other methods (see e.g. \cite{kupershmidt85}). The generalization of the $R$-matrix approach to the $q$-deformed Gelfand-Dickey structures was developed in \cite{pirozerski-semenov2000}.

Belavin and Drinfel'd in \cite{belavin-drinfeld83} obtained the $R$-matrix formulation for the finite-dimensional Toda-like equations, introduced by Bogoyavlensky \cite{bogoyavlensky76}, associated with simple Lie algebras. 
They also suggest a generalization of this construction to the two-dimensional analogue of the Bogoyavlensky systems.

However, the $R$-matrix formulation of the two-dimensional Toda hierarchy, when the number of dependent variables becomes infinite, as well as its tri-Hamiltonian structure, apparently was not considered in the literature.

Here we show how to apply the $R$-matrix approach of \cite{semenov84} - \cite{li-parmentier89} to this case. We introduce an algebra given by pairs of difference operators and then define an R-matrix associated to a non-trivial splitting of such algebra. The tri-Hamiltonian structure is then obtained by Dirac reduction to an affine subspace given by operators of fixed order (with leading coefficient equal to $1$).

This paper is organized as follows: in Section 2 the Lax representation of the two-dimensional Toda hierarchy is introduced following Ueno-Takasaki \cite{ueno-takasaki84}. In Section 3 the basic facts about the classical $R$-matrix theory applied to integrable systems are recalled. 
In Section 4 we define an algebra of difference operators which is naturally associated with the two-dimensional Toda hierarchy. 
Then we introduce a $R$-matrix and we show that it comes from a non trivial splitting of the algebra. 
We check that the skew-symmetric part of the $R$-matrix satisfies the modified Yang-Baxter equation and thus defines three compatible Poisson structures.
In Section 5, by a Dirac reduction to an affine space, we obtain three compatible Poisson structures for the two-dimensional Toda lattice hierarchy.
In Section 6 we derive the coordinate form of the Poisson brackets on the space of dependent variables.
In Section 7 we derive the tri-Hamiltonian formulation of the two-dimensional Toda flows. In the last Section we discuss some open problems.

\section{The two-dimensional Toda lattice}

The two-dimensional Toda lattice equation \cite{mikhailov79} is 
\begin{equation}
  \label{i1}
  \frac{\pa^2}{\pa \bt_1 \pa t_1} u(n) = e^{u(n) -u(n-1)} - e^{u(n+1) - u(n)} 
\end{equation}
where $u$ is a function of the continuous variables $t_1$ and $\bt_1$ and of the discrete variable $n \in \Z$.

This equation is included in the two-dimensional Toda lattice hierarchy introduced by Ueno and Takasaki \cite{ueno-takasaki84}.
We will recall the definition of this hierarchy, using the language of difference operators.

Consider two difference operators of the form
\begin{align} \label{u1.1}
  L &= \L + u_0 + u_{-1} \L^{-1} + \dots  \\
  \bar{L} &= \bar{u}_{-1} \L^{-1} + \bar{u}_0 + \bar{u}_{1} \L + \dots
\end{align}
where the coefficients are independent functions on the infinite lattice $\Z$ . $\L$ denotes the shift operator that acts on a function $f$ by $\L f(n) = f(n+1)$.

The two-dimensional Toda lattice hierarchy is given by two sets of flows, denoted by the times $t_q$ and $\bar{t}_q$ with $q>0$, by the following Lax equations
\begin{align}
L_{t_q} &= [(L^q)_+, L] & \bar{L}_{t_q} &= [(L^q)_+, \bar{L}] \label{2dtoda1} \\
\intertext{and}
L_{\bar{t}_q} &= [(\bar{L}^q)_-, L] & \bar{L}_{\bar{t}_q} &= [(\bar{L}^q)_-, \bar{L}] \label{2dtoda2} .
\end{align}

In these formulas the symbols $(\cdot)_+$ and $(\cdot)_-$ denote the projections that associate to the operator $X= \sum_k a_k \L^k$ the positive $X_+ = \sum_{k \geq 0} a_k \L^k$  and strictly negative $X_- = \sum_{k< 0} a_k \L^k$ parts, respectively.

Clearly these difference operators can be equivalently seen as infinite matrices, i.e. elements of ${gl}((\infty))$ in the formalism of \cite{ueno-takasaki84}.

All the commutators in the above equations are well-defined since the difference operators $(L^p)_+$ and $(\bL^q)_-$ are of bounded order. 
However notice that in the first part of equation \eq{2dtoda1} we can equivalently use $-(L^q)_-$ instead of $(L^q)_+$ since $L^q$ commutes with $L$, while in the second part we must use $(L^q)_+$, otherwise the commutator is not well-defined; this observation (and the analogous for the equations \eq{2dtoda2}) is useful in the construction of the $R$-matrix formulation of the hierarchy.

The flows defined above commute: this follows from the zero-curvature or Zakharov-Shabat representation given by the following 
\begin{prop} \label{u1.3}
The Toda lattice hierarchy \eq{2dtoda1}, \eq{2dtoda2} is equivalent to the system of equations
\begin{align}
& \pa_{t_p} (L^q)_+ - \pa_{t_q} (L^p)_+ + [ (L^q)_+ , (L^p)_+ ] = 0  \label{zs1}\\
& \pa_{\bt_p} (\bL^q)_- - \pa_{\bt_q} (\bL^p)_- + [ (\bL^q)_- , (\bL^p)_- ] = 0 \label{zs2} \\
& \pa_{\bt_p} (L^q)_+ - \pa_{t_q} (\bL^p)_- + [ (L^q)_+ , (\bL^p)_- ] = 0  \label{zs3}.
\end{align} 
\end{prop}

In particular, the two-dimensional Toda lattice equation \eq{i1} is obtained from equation \eq{zs3} with $p=q=1$ by the change of dependent variable $\bu_{-1}(n) = e^{u(n) - u(n-1)}$.

\pf
This has been proved in \cite{ueno-takasaki84}. For convenience of the reader we will give a slightly modified proof here, in our notations. 

Suppose the Lax equations \eq{2dtoda1}-\eq{2dtoda2} hold. Then by
\begin{align}
\pa_{t_p} \left( (L^q)_+ \right) &= \left( \pa_{t_p} L^q \right)_+ = [ (L^p)_+ , L^q ]_+ \\
&= [ (L^p)_+, (L^q)_+ ] + [ (L^p)_+, (L^q)_- ]_+ \\
&=  [ (L^p)_+, (L^q)_+ ] + \pa_{t_q} \left( (L^p)_+ \right) 
\end{align}
we obtain \eq{zs1} and similarly
\begin{align}
\pa_{\bt_p}(L^q)_+ - \pa_{t_q}(\bL^p)_- &= [ (\bL^p)_-, L^q ]_+ - [(L^q)_+, \bL^p]_- \\
&= [ (\bL^p)_- , (L^q)_+ ]_+ - [ (L^q)_+ , (\bL^p)_- ]_- \\
&= [ (\bL^p)_- , (L^q)_+ ]
\end{align}
we get \eq{zs3}. Equation \eq{zs2} is derived in the same way.

To prove the converse statement one can simply repeat the argument in \cite{ueno-takasaki84}. Rewrite
\eq{zs1} as
\begin{equation} \label{u1.5}
\pa_{t_q} L^p - [ (L^q)_+ , L^p ] = \pa_{t_q} (L^p)_- + \pa_{t_p} (L^q)_+ - [ (L^q)_+, (L^p)_- ]
\end{equation}
and observe that the RHS has order bounded by $q-1$. But if 
\begin{equation}
L_{t_q} - [(L^q)_+,L] \not= 0 
\end{equation}
then we can make the order of the LHS in \eq{u1.5} arbitrary big by increasing $p$. Hence we get (\ref{2dtoda1}a). The other Lax equations are obtained similarly.
\epf

\section{$R$-matrix and Poisson structures}

A standard method for introducing new Poisson brackets on a Lie algebra $\cG$ is given by the $R$-matrix formalism.
Here we will briefly recall some results from \cite{semenov84}, \cite{oevel-ragnisco89} and \cite{li-parmentier89}.

Let $\cG$ be a Lie algebra with a non-degenerate invariant inner product $(,)$ by which we identify it with its dual $\cG^*$.
The usual Poisson-Lie bracket on (the dual of) $\cG$ is given by
\begin{equation}
  \label{eq:ar1}
  \{ f , g \} (L) = (L , [df,dg] ) 
\end{equation}
where $f,g$ are functions on $\cG$ and the differentials $df, dg$ at the point $L \in \cG$ are identified with elements of $\cG$. 

Following \cite{semenov84} we define a $R$-matrix as a linear map $R: \cG \rightarrow \cG$ such that the bracket
\begin{equation}
  \label{eq:ar2}
  [ X, Y ]_R :=  [ R(X) , Y] + [ X, R(Y) ]  \quad  \text{for} \quad X,Y \in \mathcal{G}
\end{equation}
satisfies the Jacobi identity.

Hence given a $R$-matrix we have another linear Poisson bracket on $\cG$, given by the Poisson-Lie bracket associated with $[,]_R$
\begin{equation}
  \label{eq:ar3}
  \{ f , g \}_1 (L) = (L , [df,dg]_R ) .
\end{equation}

A sufficient condition for $R \in \End(\mathcal{G})$ to be a $R$-matrix is that 
\begin{equation} \label{mYBeq}
[R(X),R(Y)]-R([X,Y]_R)= - [X,Y] .
\end{equation}
Equation \eq{mYBeq} is called {\it modified Yang--Baxter equation.}

As pointed out in \cite{semenov84}, the so-called Sklyanin brackets, which are quadratic in $L$, can be defined on an associative algebra $\cG$ if the $R$-matrix is skew-symmetric and satisfies the modified Yang--Baxter equation.
This result has been generalized for the non-unitary case (when the $R$-matrix is not skew-symmetric with respect to the inner product) by Oevel--Ragnisco \cite{oevel-ragnisco89} and Li--Parmentier \cite{li-parmentier89}. 

Let $\cG$ be an associative algebra, with the natural Lie bracket given by the commutator. Assume on $\cG$ the existence of a symmetric trace-form $\tr: \cG \to \C$ with the associated non-degenerate invariant inner product
\begin{equation}
  (L_1 , L_2 ) := \tr(L_1 L_2) ; 
\end{equation}
using this inner product we identify $\cG$ and $\cG^*$.
Let $R: \cG \to \cG$ be a linear map. Define the following three brackets on the functions on $(\cG)$
\begin{subequations}
  \label{genpoibra}
  \begin{align}
    \{ f_1, f_2 \}_1  &:= \frac12 (L, [df_1, df_2 ]_R ) = \frac12 ([L,df_1], R(df_2) ) - \frac12 ([L,df_2], R (df_1)) \\
    \{ f_1, f_2 \}_2  &:= \frac14 ([L,df_1], R(L df_2 + df_2 L))- \frac14 ([L,df_2], R(Ldf_1 + df_1 L)) \\
    \{ f_1, f_2 \}_3  &:= \frac12([L,df_1], R(L df_2 L)) -\frac12 ([L,df_2], R(L df_1 L)) .
  \end{align}
\end{subequations}
The first bracket is simply the linear Poisson-Lie bracket associated to the $R$-matrix that we have considered above, rescaled by a factor $\frac12$.
We have 
\begin{prop}[\cite{oevel-ragnisco89, li-parmentier89}]  
\label{teo-ragnisco}
(1) For any $R$-matrix $R$, $\{,\}_1$ is a Poisson bracket. \\
(2) If both $R$ and its skew-symmetric part $A = \frac12 (R- R^*)$ satisfy the modified Yang-Baxter equation \eq{mYBeq} then $\{ , \}_2$ is a Poisson bracket. \\
(3) If $R$ solves the modified Yang-Baxter equation \eq{mYBeq} then $\{, \}_3$ is a Poisson bracket. \\
Moreover the three brackets are compatible.
\end{prop}

We recall that two Poisson brackets are said to be {\it compatible} if any their linear combination is still a Poisson bracket.

The  Poisson tensors $P_i$ corresponding to the brackets \eq{genpoibra} are defined by
\begin{equation}
\label{poiss-bra-tens} 
\{ f_1 ,f_2 \}_i (L) =: ( df_1 , P_i(L) df_2 ) \quad i=1,2,3
\end{equation}
and are explicitly given by
\begin{eqnarray}
P_1(L) df &=& \frac12  [ R(df), L]- \frac12 R^* ([L, df])  \\
P_2(L) df &=& \frac14 [ R(L df + df L), L] - \frac14 L R^*([L,df]) -\frac14  R^*([L,df])L \\
P_3(L) df &=& \frac12 [ R(L df L ), L] -\frac12 L R^*([L,df])L .
\end{eqnarray}

The Hamiltonian flow associated by the Poisson bracket $\{ , \}_i$ to the Hamiltonian function $H$ on $\cG$ is defined by the usual formulas
\begin{equation}
  \label{eq:ar4}
  f_{t_i} := \{ f, H \}_i .  
\end{equation}
It is clear that the corresponding vector fields are given by
\begin{equation}
  \label{eq:ar5}
  L_{t_i} = P_i dH  \quad \text{for} \quad  i=1,2,3. 
\end{equation}

Among the possible Hamiltonian functions on $\cG$ a particular role is played by the Casimirs of the Poisson bracket $\{,\}$ defined in \eq{eq:ar1}, i.e.  functions $H$ on $\cG$ such that $\{ H, f \}=0$ for any function $f$ on $\cG$. 
\begin{prop}[\cite{oevel-ragnisco89, li-parmentier89}] \label{pro4}
  The Casimir functions of $\{,\}$ are in involution with respect to the three Poisson brackets \eq{genpoibra}. If $H$ is a Casimir of $\{,\}$ then the associated Hamilton equations \eq{eq:ar5} have the following Lax form
  \begin{align}
    L_{t_1} &= [ \frac12 R(dH), L ] \label{q348}\\
    L_{t_2} &= [ \frac14 R( L dH + dH L) , L ] \\
    L_{t_3} &= [ \frac12 R(L dH L), L ]  .
  \end{align}
\end{prop}

Later we will need to reduce on a certain submanifold the Poisson brackets defined on the whole algebra, following the general Dirac prescription. Since we will always deal with reductions to affine subspaces we recall the following well-known
\begin{lemma}[see e.g. \cite{oevel-ragnisco89}] \label{lemma-reduction}
Given two linear spaces $U$, $V$ with coordinates $u$, $v$, let 
\begin{equation}
  P(u,v) = 
  \begin{pmatrix} 
    P_{uu} & P_{uv} \\ P_{vu} & P_{vv}
  \end{pmatrix} 
  : U^* \oplus V^* \to U \oplus V
\end{equation}
be a Poisson tensor on $U \oplus V$. If the component $P_{vv} : V^* \to V$ is invertible then, for an arbitrary $v \in V$, the map $P^{red} (u;v): U^* \to U$ given by 
\begin{equation} 
  \label{dirac.1}
  P^{red} (u; v) =P_{uu}(u,v) - P_{uv} (u,v) (P_{vv}(u,v))^{-1} P_{vu}(u,v)
\end{equation}
is a Poisson tensor on the affine space $v+U \subset U\oplus V$.
\end{lemma}

The condition of invertibility of $P_{vv}$ can be relaxed by asking that the images and the kernels of the different components satisfy
\begin{align}
  \label{star}
 &\im P_{vu} \subset \im P_{vv} \\
 &\Ker P_{vv} \subset \Ker P_{uv},
\end{align}
where $P_{uv}:V^*\rightarrow U$ and $P_{vu}:U^* \rightarrow V$.

\section{Algebras of difference operators and $R$-matrix}
We denote by $\cF$ the space of functions on the lattice $\Z$ and by $\cA$ the space of (formal) difference operators with coefficients in $\cF$
\begin{equation}
  \label{l1}
  \cA = \left\{ \sum_{k \in \Z} a_k(n) \L^k \right\} .
\end{equation}

The subspaces $\cA^{\pm}$ of $\cA$ are given by operators that have respectively upper or lower bounded order
\begin{align}
  \label{l2}
  &\cA^+ = \left\{ \sum_{k \leq K} a_k(n) \L^k  \quad \text{for some} \ K \in \Z \right\}  \\
  \label{l3}
  &\cA^- = \left\{ \sum_{k \geq K} a_k(n) \L^k  \quad \text{for some} \ K \in \Z \right\}  ,
\end{align}
while $\cA^0 = \cA^+ \cap \cA^-$ is the space of operators that have both upper and lower bounded order. 

The spaces $\cA^+$, $\cA^-$ and $\cA^0$ are associative algebras with the usual multiplication defined by $\L f (n) = f(n+1) \L$. Hence they are Lie algebras with the Lie bracket given by the commutator. On the other hand the product is not well defined in $\cA$ due to the presence of infinite sums.

For each $k \in \Z$ we define on these spaces the projections $(\cdot )_{\geq k}$ and $(\cdot )_{\leq k}$ that associate to the difference operator $X = \sum_{l\in \Z} a_l \L^l$ the operators
\begin{equation}
  \label{l4}
  X_{\geq k} = \sum_{l \geq k} a_l \L^l \qquad X_{\leq k} =  \sum_{l \leq k} a_l \L^l .
\end{equation}

We will also use the subscripts $+$ and $-$ with the usual meaning
\begin{equation}
  \label{l5}
  X_+ = X_{\geq 0} \qquad X_{-} = X_{< 0} .
\end{equation}

A symbol like $(\cA^+)_{\leq 0}$ will denote the obvious space of difference operators obtained by projection.

The residue of $X= \sum_{k \in \Z} a_k \L^k$   is defined by
\begin{equation}
\res X := a_0 .
\end{equation}
It is easy to verify that the residue of a commutator is always given by the discrete derivative of a function
\begin{equation}
  \label{l6}
  \res [X, Y](n) =  f(n+1) - f(n) \quad \text{for some} \quad f \in \cF .
\end{equation}
From translation invariance of the sum $\sum_{n\in \Z} f(n)$ it follows that
$\tr [X,Y] = 0$, where the trace-form of a difference operator is defined by
\begin{equation} \label{l7}
\tr X := \sum_{n \in \Z} \res X(n)   .
\end{equation}
The bilinear pairing 
\begin{equation} \label{x.1}
(X,Y) := \tr XY
\end{equation}
gives a non-degenerate invariant symmetric inner product on $\cA^+$, $\cA^-$ and $\cA^0$.

It is understood that one should actually restrict to a subspace of $\cF$ given by functions such that the previous summations make sense. As is usually done in formal discussions like this we will leave such conditions unspecified. 

We will need to consider the duals to some of the spaces defined above.
To avoid any topological complication we will simply define
\begin{equation}
  \label{l9}
  (\cA^+)^* = \cA^+ \qquad (\cA^-)^* = \cA^- 
\end{equation}
in the sense that we will consider only the linear functionals on $\cA^{\pm}$ induced by elements of $\cA^{\pm}$ through the inner product.
The dual to $(\cA^+)_{\geq k}$ is then clearly given by $\cA^+$ up to elements of $(\cA^+)_{\geq -k+1}$ and it can be identified with $(\cA^+)_{\leq -k}$ ; in general
\begin{equation}
  \label{l10}
  ((\cA^{\pm})_{\geq k})^* = (\cA^{\pm})_{\leq -k} \qquad 
  ((\cA^{\pm})_{\leq k})^* = (\cA^{\pm})_{\geq -k} .
\end{equation}

Let us define the associative algebra 
\begin{equation}
   \mathfrak{A} := \cA^+ \oplus \cA^- .
\end{equation}
Elements of $\mathfrak{A}$ will be written as pairs of difference operators $(X,\bX) \in \mathfrak{A}$ where $X \in \cA^+$, $\bX \in \cA^-$.

The inner product on $\mathfrak{A}$ is defined by the trace form
\begin{equation}
  \label{u2.1}
  \tr X \oplus \bX = \tr X + \tr \bX
\end{equation}
for $X \oplus \bX \in \mathfrak{A}$.

Define the $R$-matrix, $R: \mathfrak{A} \rightarrow \mathfrak{A}$ by 
\begin{equation}
  \label{u2.2}
  R(X,\bX) = (X_+ - X_- + 2 \bX_- , \bX_- - \bX_+ +2 X_+) 
\end{equation}
where $(X , \bX ) \in \mathfrak{A}$.

We emphasize that this $R$-matrix is not simply given by a direct sum of the usual $R$-matrices on $\cA^+$ and $\cA^-$ associated to the splitting of these algebras in positive and negative parts; this is due to the fact that the second equation in \eq{2dtoda1} and the first equation in \eq{2dtoda2} give a coupling between $L$ and $\bL$, that must be taken into account.

\begin{thm}
The linear maps $R$ and $A=\frac12 (R-R^*)$ satisfy the modified Yang-Baxter equation \eq{mYBeq}.
\end{thm}
\pf 
$R$ is given by a splitting of the Lie algebra $\mathfrak{A}$ hence it satisfies \eq{mYBeq}. Indeed we can write
\begin{equation}\label{u2.3}
R = \Pi - \tPi
\end{equation}
where $\Pi$ and $\tPi$, defined by
\begin{equation}\label{u2.4}
  \Pi(X, \bX) = (X_+ + \bX_- , X_+ + \bX_-) \qquad 
  \tPi (X, \bX) = ( X_- - \bX_- , \bX_+ - X_+ ) ,
\end{equation}
are projections operators, i.e. $\Pi^2 = \Pi$, $\tPi^2 = \tPi$, $\tPi \Pi = 0 = \Pi \tPi $ and $\Pi + \tPi = Id$. Thus the splitting\footnote{Such splitting of the algebra $\mathfrak{A}$ has already been used in \cite{vanmoerbeke-mumford79} and \cite{adler-vanmoerbeke80}. However, to our best knowledge, the associated $R$-matrix has not been considered in the literature. } associated to the $R$-matrix above is
\begin{equation}\label{u2.5}
\mathfrak{A} = \Big( \diag ( \cA^0 \oplus \cA^0 ) \Big) \oplus 
\Big(  (\cA^+)_- \oplus (\cA^-)_+     \Big) .
\end{equation}

As in the ``one dimensional'' case (given by the algebra $\cA^+$ with the splitting $(\cA^+)_+ \oplus (\cA^+)_-$) this splitting is not isotropic with respect to the inner product. Indeed the $R$-matrix is not skew-symmetric, the adjoint $R^*$ being given by
\begin{equation} \label{u2.6}
  R^*(X, \bX) = (X_{\leq 0} - X_{>0} + 2 \bX_{\leq0} , \bX_{>0} - \bX_{\leq0} + 2 X_{>0} ) = \Pi^* - \tPi^*
\end{equation} 
where the dual projections are 
\begin{align}\label{u2.7}
  &\Pi^*(X, \bX) = (X_{\leq 0} + \bX_{\leq 0} , X_{>0} + \bX_{>0} ) \\
  &\tPi^*(X, \bX) = (X_{>0} - \bX_{\leq 0}, \bX_{\leq 0} - X_{>0} ) .
\end{align} 

By substitution in equations \eq{eq:ar2} and \eq{mYBeq} one can also verify that
the skew-symmetric part $A$ of the $R$-matrix \eq{u2.2} 
\begin{equation}\label{u2.8}
  A(X , \bX) = ( X_{>0} -X_{<0} - \bX_0 ,
  \bX_{<0} - \bX_{>0} + X_0 )
\end{equation} 
satisfies the modified Yang-Baxter equation \eq{mYBeq}.
\epf

Thus, by the previous general theorems, there are on $\mathfrak{A}$ three compatible Hamiltonian structures \eq{genpoibra}.
We summarize this result and the explicit form of the Poisson tensors in the following
\begin{thm}
  On the algebra $\mathfrak{A}$ there are three compatible Poisson structures given by
  \begin{subequations} \label{eq:ar10}
    \begin{align}
      P_1( X , \bX) &= \Big( [L, X_- - \bX_-] - ([L,X] + [\bL,\bX])_{\leq 0} , \\
      & [\bL, \bX_+ - X_+ ] - ([L,X] + [\bL, \bX])_{>0} \Big) \notag\\
      P_2( X , \bX) &= \Big( \frac12 [L, (L X + X L)_- - (\bL \bX +\bX \bL)_-]  \\ 
      &- \frac12 L( [L, X]_{\leq0} + [\bL, \bX]_{\leq0}) - \frac12( [L, X]_{\leq0} + [\bL, \bX]_{\leq0})L  , \notag \\
      &\frac12 [\bL,(\bL \bX + \bX \bL)_+ -(L X + X L)_+ ] \notag \\
      &- \frac12 \bL ([L, X]_{>0} +[\bL, \bX]_{>0} )  - \frac12([L, X]_{>0} +[\bL, \bX]_{>0} ) \bL \Big) . \notag \\
  P_3( X, \bX ) &= \Big(  [ L, (LXL - \bL \bX \bL)_- ] -  L \big( [L,X] + [ \bL, \bX ] \big)_{\leq 0} L , \\
  & [ \bL , (\bL \bX \bL - LXL)_+ ] - \bL \big( [ \bL , \bX ] + [ L, X] \big)_{>0} \bL \Big) \notag 
    \end{align}
  \end{subequations}
\end{thm}
It is understood that these are the Poisson tensors at the point $(L,\bL) \in \mathfrak{A}$ and they are evaluated on the co-vector $(X, \bX) \in \mathfrak{A}^* =  \mathfrak{A} $.

\section{A Poisson pencil for the two-dimensional Toda hierarchy}

To calculate the Poisson brackets for the two-dimensional Toda hierarchy we need to reduce the Poisson tensors, found so far on the algebra $\mathfrak{A}$, to the affine subspace $\L \oplus 0 + (\cA^+)_{\leq 0} \oplus (\cA^-)_{\geq -1}$ given by pairs $(L,\bL)$ of operators of the form
\begin{equation} \label{aa.1}
L= \L + u_0 + \dots  \qquad \bar{L}= \bu_{-1} \L^{-1} + \bu_0 +\dots .
\end{equation} 

We will assume $(L, \bL) \in (\L , 0) + U$ where, following the notations of Lemma \ref{lemma-reduction}
\begin{equation}
  \label{uv1}
  U := (\cA^+)_{\leq 0} \oplus (\cA^-)_{\geq -1} \quad V := (\cA^+)_{\geq 1} \oplus (\cA^-)_{\leq -2} ,
\end{equation}
and the dual spaces are
\begin{equation}
  \label{uv2}
  U^* = (\cA^+)_{\geq 0} \oplus (\cA^-)_{\leq 1} \quad V^* = (\cA^+)_{\leq -1} \oplus (\cA^-)_{\geq 2} .
\end{equation}

The component $P_{vu}^{(1)} : U^* \rightarrow V$ of $P_1$ is easily seen to be zero. Hence the correction term in \eq{dirac.1}
 is zero and the reduced Poisson tensor $P_1^{red}$ is simply given by the restriction of (\ref{eq:ar10}a)
\begin{equation}
  \label{uv3}
  P_1^{red} (X, \bX ) = \Pi_U P_1 (X, \bX) = P_1 (X,\bX) , 
\end{equation}
where $\Pi_U$ indicates the projection on the subspace $U$, parallel to $V$.

Notice that also the components $P_{uv}^{(1)}$ and $P_{vv}^{(1)}$ of $P_1$ are zero for $(L,\bL) \in (\L,0) + U$.

The component $P_{vu}^{(2)}$ of the second Poisson tensor (\ref{eq:ar10}b) is however non zero
\begin{equation}
  \label{uv4}
  P_{vu}^{(2)}(X,\bX) = \Pi_V P_2(X, \bX) = - \frac12 ( \L \res ([L,X]+[\bL,\bX]) + \res ([L,X]+[\bL,\bX]) \L , 0 ) ,
\end{equation}
where $(X, \bX) \in U^*$. 
The image of this operator is given by elements of the form 
\begin{equation}
  \label{uv5}
  (a \L, 0) \quad \text{for} \quad a \in (\L -1) \cF .
\end{equation}
It turns out that conditions \eq{star} are satisfied, i.e. we can unambiguously invert $P_{vv}^{(2)}$ on such image. We obtain that the correction term in \eq{dirac.1} is given by 
\begin{equation}
  \label{eq:sp2}
  -P_{uv}^{(2)} \circ (P_{vv}^{(2)})^{-1} \circ P_{vu}^{(2)} (X,\bX) = \Big( 
  \frac12 \left[ L_{\leq0} , \zeta \right] , 
  \frac12 \left[ \bL       , \zeta \right] 
  \Big) ;
\end{equation}
$\zeta \in \cF$ is given by 
\begin{equation}
  \label{uv6}
  \zeta =  (\L+1) (\L-1)^{-1} \res ([L,X] + [\bL , \bX] ) 
\end{equation}
where the shift operators act only on the function given by the residue.

Even if the inverse operator $(\L-1)^{-1}$ appears in the previous equation, it is applied on the residue of a commutator i.e. on elements of the form \eq{uv5}, hence it always cancels.

From \eq{dirac.1} one finds that the reduced second Poisson tensor is given by
\begin{align} \label{aa.2}
  P_2^{red}(X , \bX) &= \Big( \frac12[ L, (L X + X L)_- - ( \bL \bX + \bX \bL )_- ] +\frac12 [L,\zeta ] \\
  & -\frac12 L( [ L , X ] +  [ \bL, \bX])_{\leq 0}  -\frac12 ( [ L, X ] + [ \bL, \bX ] )_{\leq 0} L 
   , \notag \\
  & \frac12 [ \bL, (\bL \bX + \bX \bL)_+  - (L X + X L)_+ ] + \frac12 [ \bL, \zeta ]   \notag \\
  & -\frac12 \bL ( [ L, X ] + [ \bL, \bX ] )_{>0} -\frac12 ([ L, X ] + [\bL , \bX ] )_{>0} \bL 
  \Big) \notag,
\end{align}
with $\zeta$ defined in \eq{uv6}.

We find that the third Poisson tensor also needs a correction term. After some computations 
we obtain the following Dirac reduced Poisson tensor
\begin{align}
  \label{e1}
  P^{red}_3(X,\bX) = &\Big(
  [ L, (LXL - \bL \bX \bL)_- ] 
  - L ( [L,X] +[\bL, \bX] )_{\leq -2} L  
  -( L ( [L,X] + [\bL, \bX] )_{0,-1} L )_{\leq 0} 
  +[L , \mathcal{Z}]_{\leq 0} , \notag \\
  &[ \bL, (\bL \bX \bL - LXL)_+] 
  -  \bL  ( [\bL,\bX] + [L,X])_{>0} \bL
  + [\bL, \mathcal{Z} ]_{\geq -1} 
  \Big) 
\end{align}
where
\begin{equation}
  \label{e2}
  \mathcal{Z} = \alpha \L +\beta
\end{equation}
and the functions $\alpha$, $\beta$ and $\gamma$ are given by
\begin{align}
  \label{e3}
  &\alpha = \L (\L-1)^{-1} \res ( [L,X] + [ \bL, \bX ]   ) \\
  &\beta = (\L -1)^{-1} \Big[ 
  (\L u_0) (\L \alpha) -u_0 (\L^{-1} \alpha) +\L \gamma 
  \Big] \\
  &\gamma \L^{-1} = ([L,X] + [\bL, \bX])_{-1} .
\end{align}

We have thus proved part of the following
\begin{thm}
  The maps 
  \begin{equation}
    P_i^{red} : \left( (\cA^+)_{\leq 0} \oplus (\cA^-)_{\geq-1} \right)^* = (\cA^+)_{\geq 0} \oplus (\cA^-)_{\leq1} 
    \rightarrow (\cA^+)_{\leq 0} \oplus (\cA^-)_{\geq -1}
  \end{equation}
for $i=1,2,3$, defined by formulas (\ref{eq:ar10}a), \eq{aa.2} and \eq{e1}, are three compatible Poisson tensors on the affine space of operators $L$ and $\bL$ of the form \eq{aa.1}.
\end{thm}

\pf
We have to show that the three reduced Poisson tensors are compatible.

As observed above, for $(L, \bL)$ in the affine space under consideration, the only non zero component of $P_1$ is $P_{uu}^{(1)}$; from this easily follows that the reduction $P_{\lambda}^{red}$ of the Poisson pencil $P_{\lambda} := P_1 + \lambda P_2$ is given by the pencil of the reduced Poisson tensors $P_1^{red} + \lambda P_2^{red}$, hence the reduced Poisson tensors are compatible.

For the same reason also $P^{red}_1$ and $P^{red}_3$ are compatible Poisson tensors, thus we just need to prove the compatibility of $P_2$ and $P_3$. 
This follows from the following observation.

Define on the algebra $\mathfrak{A}$ the following constant vector field
\begin{equation}
  \label{z1}
  \phi (L, \bL) = (1,1)
\end{equation}
that generates the flow
\begin{equation}
  \label{z2}
  \Phi^t (L,\bL) = (L+t , \bL +t) .
\end{equation}

The push-forward of the Poisson tensors is simply given by
\begin{equation}
  \label{z3}
  \Phi^t_* P_i (L, \bL) = P_i (L-t , \bL -t ) .
\end{equation}

In particular the action on the Poisson tensors \eq{eq:ar10} is given by
\begin{subequations}
  \label{z4}
  \begin{align}
    &\Phi^t_* P_1 = P_1 \\
    &\Phi^t_* P_2 = P_2 -t P_1 \\
    &\Phi^t_* P_3 = P_3 -2t P_2 + t^2 P_1 .
  \end{align}
\end{subequations}

The vector field $\phi$ and the associated flow $\Phi^t$ are well defined even on the affine subspace $(\L,0) + \mathfrak{A}$. Moreover one can check that the reduced Poisson tensors $P_i^{red}$, given in \eq{aa.2} and \eq{e1}, transform under exactly the same formulas \eq{z4}. 

Since the push-forward preserves the Poisson properties of a tensor, from the second formula we immediately deduce that $P^{red}_1$ and $P^{red}_2$ are compatible, as already observed above. From the third formula, using the bilinearity of the Schouten bracket and the fact that $P^{red}_1$ is compatible with $P^{red}_2$ and $P^{red}_3$ we easily obtain the compatibility of $P^{red}_2$ and $P^{red}_3$.
\epf

\section{Poisson brackets}
We want to express the Poisson brackets associated to $P_k^{red}$ in the usual coordinate form  $\{ u_i(n), u_j(m) \}_{k}$. Essentially we have to evaluate \eq{poiss-bra-tens} on functions 
of the forms
\begin{equation}
  \label{eq:ar7}
  f(L,\bL) = u_i(m) \quad \text{and} \quad f(L,\bL) = \bu_j(m) ,
\end{equation}
where $i\leq 0$ , $j \geq -1$ and $m \in \Z$.
The associated differentials are 
\begin{equation}
  \label{eq:ar8}
  df = (\L^{-i} \de (n-m) , 0 ) \quad \text{and} \quad df = (0,  \L^{-j} \de (n-m) ) 
\end{equation}
respectively.

By substituting these formulas and the Poisson tensors (\ref{eq:ar10}a), \eq{aa.2} and \eq{e1} in \eq{poiss-bra-tens} we obtain the following expressions of the Poisson brackets.

\noindent
\underline{First bracket}
\begin{subequations}
\label{first2dexplbra}
\begin{align}
  \{ u_i(n) , u_j(m) \}_1  &=   u_{i+j}(m) \de(n-m-j)- u_{i+j}(n) \de(n-m+i)  \\
  \{ u_i(n) , \bu_j(m) \}_1  &=  c(j) [ u_{i+j}(m) \de(n-m-j) - u_{i+j}(n) \de(n-m+i) ]  \\
  &+  [   \bu_{i+j}(m) \de(n-m-j) - \bu_{i+j}(n) \de(n-m+i) ] \notag \\
  \{ \bu_i(n) , \bu_j(m) \}_1 &=  (1-c(i)-c(j)) [\bu_{i+j}(n) \de(n-m+i) -  \bu_{i+j}(m) \de(n-m-j) ] ;
\end{align}
\end{subequations} 
the function $c(i)$ is defined to be $1$ for $i>0$ and $0$ otherwise.

We have assumed that whenever the variable $u_k$ (or $\bu_k$) appears on the RHS of \eq{first2dexplbra}, it is regarded as having value zero if $k>1$ (or $k < -1$ respectively); for example the RHS of (\ref{first2dexplbra}a) is non-zero only if $i+j \leq 1$.  Moreover recall that $u_1=1$. 

\noindent
\underline{Second bracket}
{\Small
\begin{subequations} \label{d2pb}
  \begin{align}
    \{ u_i(n) , & u_j(m) \}_2 = 
    u_i(n) u_j(m) \sum_{s=i}^0 \big[ \de(n-m+s-j) - \de(n-m+s) \big] \\
    &+\sum_{s=1}^{1-i} \big[ 
    u_{i+s}(n) u_{j-s}(m) \de(n-m+i-j+s) 
    -u_{j-s}(n) u_{i+s}(m) \de(n-m-s) \big] \notag \\
    \{ u_i(n) , & \bu_j(m) \}_2 =
    u_i(n) \bu_j(m) \sum_{s=i}^0 \big[ 
    \de(n-m+s-j) - \de(n-m+s) \big] \\
    &+ \sum_{s=1}^{\min(1+j,1-i)} \big[ 
    u_{i+s}(n-s) \bu_{j-s}(m) \de(n-m-j) 
    -u_{i+s}(n) \bu_{j-s}(m+s) \de(n-m+i) \big] \notag \\
    \{ \bu_i(n) , & \bu_j(m) \}_2 = 
    \bu_i(n) \bu_j(m) \sum_{\substack{s=-j \\ j\not= -1}}^0 \big[
    \de(n-m+s+i) - \de(n-m+s) \big] \\
    &+\sum_{s=1}^{i+1} \big[
    \bu_{i-s}(n) \bu_{j+s}(m) \de(n-m+i-j-s) 
    - \bu_{j+s}(n) \bu_{i-s}(m) \de(n-m+s) \big] \notag
  \end{align}
\end{subequations}
}
Observe that the first sum in (\ref{d2pb}c) is zero for $j=-1$.

\noindent
\underline{Third bracket}
\begin{subequations}
  \label{ee5}
  {\Small
    \begin{equation}
      \begin{split}
        &\{ u_i(n) , u_j(m) \}_3 = \\ 
        &\sum_{\substack{i+1 \leq s \leq 1 \\ k \leq 1 \\ k+s \geq i+j-1}}
        \big[ u_s(n) u_k(n+s) u_{i+j-k-s}(m) \de(n-m+k+s-j) - u_{i+j-k-s}(n) u_s (n+i-s) u_k(m) \de(n-m+i-k-s)  \big]   \\
        &- \sum_{\substack{k \leq j \\ s \leq 1 \\ k+s \geq i+j-1 }}
        \big[ u_s(n) u_k(n+s) u_{i+j-k-s}(m) \de(n-m+k+s-j) - u_{i+j-k-s}(n) u_s (n+i-s) u_k(m) \de(n-m+i-k-s)  \big]  \\
        &+u_j(m) \big( u_{i-1}(n+1) \cD^{-j}[\de(n-m+1)] - u_{i-1}(n) \cD^{-j} [\de(n-m+i)] \big) \\ 
        &+u_i(n) \big( u_{j-1}(m+1) \cD^{i}[\de(n-m)] - u_{j-1}(m) \cD^i [\de(n-m-j+1)] \big) \\
        &-u_i(n) u_j(m) \cD^i \big[u_0(n+1) \cD^{-j}[ \de(n-m+2)] -u_0(n) \cD^{-j}[\de(n-m)] \big] 
      \end{split}
    \end{equation} 
    }
  
  {\Small
    \begin{align}
      &\{u_i(n) , \bu_j(m) \}_3 = \\
      &\sum_{\substack{k+1 \leq l \leq 1 \\k\geq -1 \\k+l \leq i+j+1 }}
      \big[ \bu_k(n) u_l(n+i-l) \bu_{i+j-l-k}(m) \de(n-m+k-j)
      -u_l(n) \bu_k(n+l) \bu_{i+j-k-l}(m) \de(n-m+l+k-j) \big] \notag  \\
      &-\sum_{\substack{-1 \leq l \leq j \\ k \leq 1 \\ k+l \geq j+i-1}} 
      \big[u_k(n) \bu_l(n+k) u_{j+i-k-l}(m) \de(n-m+k+l-j) 
      - u_k(n) u_{j+i-k-l}(n+k-j+l) \bu_l(m) \de(n-m+k-j) \big]  \notag \\
      &-\bu_j(m) \Big( u_{i-1}(n) \cD^{-j} [ \de(n-m+i) ] - u_{i-1}(n+1) \cD^{-j} [ \de(n-m+1)] \Big) \notag \\
      &+ u_i(n) \Big( \bu_{j-1}(m+1) \cD^i[\de(n-m)] -\bu_{j-1}(m) \cD^i [\de(n-m-j+1)] \Big) \notag  \\
      &-u_i(n) \bu_j(m) \Big( \cD^i [ u_0(n+1) \cD^{-j} [ \de(n-m+2) ] ]
      - \cD^i [ u_0(n) \cD^{-j} [\de(n-m)] ] \Big)  \notag
    \end{align} 
    }
  
  {\Small
    \begin{align}
      &\{\bu_i(n) , \bu_j(m) \}_3 = \\
      &\sum_{\substack{-1\leq k \leq i \\ s \geq -1 \\ k+s \leq i+j+1}}
      \big[ \bu_k(n) \bu_{i+j-k-s}(n+k) \bu_s(m) \de(n-m+i-s)
      -\bu_{i+j-k-s}(n) \bu_k(n+i-k) \bu_s(m) \de(n-m+i-k-s) \big]  \notag \\
      &-\sum_{\substack{l \geq j+1 \\ s \geq -1 \\ l+s \leq i+j+1}}
      \big[ \bu_{i+j-l-s} (n) \bu_l(n+i+j-l-s) \bu_s(m) \de(n-m+i-s)
      -\bu_{i+j-l-s}(n) \bu_s(n+i-s) \bu_l(m) \de(n-m+i-l-s) \big] \notag \\
      &- \bu_i(n) \bu_j(m) \cD^i \big[ u_0(n+1) \cD^{-j} [ \de(n-m+2) ]
      -u_0(n) \cD^{-j} [\de(n-m)] \big] \notag \\
      &- \bu_j (m) \big( \bu_{i-1}(n) \cD^{-j}[\de(n-m+i)] - \bu_{i-1}(n+1) \cD^{-j} [\de(n-m+1) ] \big) \notag \\
      &+ \bu_i(n) \big( \bu_{j-1} (m+1) \cD^i [\de(n-m)] - \bu_{j-1}(m) \cD^i [ \de(n-m-j+1) ] \big) \notag
    \end{align} 
    }
\end{subequations}

The operator $\cD^k$ acts on functions of $n$ in the following way
\begin{equation}
  \label{e5}
  \cD^k [f(n)] := (\L^k -1)(\L-1)^{-1}[f(n)] = 
  \begin{cases}
    \sum_{i=0}^{k-1} f(n+i)\qquad &k>0 \\
    0  &k=0 \\
    -\sum_{i=k}^{-1} f(n+i) &k<0 .
  \end{cases}
\end{equation}

We summarize this result in the following 
\begin{thm} \label{thm61}
  The brackets \eq{first2dexplbra}, \eq{d2pb} and \eq{ee5} give three compatible Poisson structures in the variables $u_i(n)$ for $i\leq0$ and $\bu_j(m)$ for $j\geq -1$ and $n,m \in \Z$.
\end{thm}
We will call these the {\it Poisson brackets for the two-dimensional Toda 
hierarchy}.

\section{Hamiltonian representation}

We define the Hamiltonian functions, for $p \geq0$
\begin{equation} \label{qxxx}
  h_p = \frac1{p+1} \tr  L^{p+1} \qquad
  \bh_p =  \frac1{p+1} \tr \bar{L}^{p+1}
\end{equation} 
where $(L, \bar{L} )$ is a point in $\mathfrak{A}$. 

These induce, through the first Poisson bracket (obtained from Poisson tensor (\ref{eq:ar10}a)), two sets of flows on $\mathfrak{A}$
\begin{equation}
  \label{ha1}
  \frac{\pa}{\pa t_p} \cdot = \{ \cdot , h_p \}_1 \qquad
  \frac{\pa}{\pa \bt_p} \cdot = \{ \cdot , \bh_p \}_1 ; 
\end{equation}
the flows $t_0$ and $\bt_0$ are trivial, since the corresponding Hamiltonians $h_0$, $\bh_0$ are Casimirs of the first bracket.

The Hamiltonians $h_p$, $\bh_p$ are Casimirs of the standard Poisson-Lie bracket \eq{eq:ar1} on (the dual of) $\mathfrak{A}$. Hence by Proposition \ref{pro4} these flows have the following Lax representation 
\begin{subequations}\label{ha2}
  \begin{align}
    L_{t_q} &= [(L^q)_+, L] & \bar{L}_{t_q} &= [(L^q)_+, \bar{L}]  \\
    L_{\bar{t}_q} &= [(\bar{L}^q)_-, L] & \bar{L}_{\bar{t}_q} &= [(\bar{L}^q)_-, \bar{L}] .
  \end{align}
\end{subequations}
Comparing the Lax representations associated by the same Proposition to the three different Poisson brackets, we observe that $h_p$, $\bh_p$ satisfy the tri-Hamiltonian recursion relation
\begin{subequations}
  \label{ha3}
  \begin{align}
    \{ \cdot , h_p \}_1 = \{ \cdot , h_{p-1} \}_2 = \{ \cdot , h_{p-2} \}_3 \\
    \{ \cdot , \bh_p \}_1 = \{ \cdot , \bh_{p-1} \}_2 = \{ \cdot , \bh_{p-2} \}_3
  \end{align}
\end{subequations}
for $p \geq 2$, with the un-reduced Poisson brackets obtained from tensors \eq{eq:ar10}; moreover 
\begin{subequations}
  \label{ha3.1}
  \begin{align}
    \{ \cdot , h_1 \}_1 = \{ \cdot , h_{0} \}_2 \\
    \{ \cdot , \bh_1 \}_1 = \{ \cdot , \bh_{0} \}_2  .
  \end{align}
\end{subequations}

Analogous relations hold in the two-dimensional Toda case. It is obvious that the Lax representation \eq{ha2} still follows from the definition of the flows \eq{ha1}, since the reduced first Poisson tensor is simply given by restriction.

Recall that, by definition of the Dirac reduction of a Poisson bracket, the restriction of functions to the affine subspace $\L \oplus 0 + \mathfrak{A}$ is a homomorphism from the Poisson algebra of functions on $\mathfrak{A}$ to the Poisson algebra of functions on the affine subspace, with the corresponding reduced Poisson bracket.

This implies that the tri-Hamiltonian recursion relation that follows from \eq{ha3} and \eq{ha3.1} holds in the two-dimensional Toda case.

In the following Theorem we summarize the tri-Hamiltonian formulation of the two-dimensional Toda lattice hierarchy.
\begin{thm}
  The two-dimensional Toda lattice flows defined by \eq{2dtoda1} and \eq{2dtoda2} admit the tri-Hamiltonian representation
  \begin{align}
    &\frac{\pa}{\pa t_p} \cdot = \{ \cdot, h_p \}_1 = \{ \cdot , h_{p-1} \}_2 = \{ \cdot , h_{p-2} \}_3 \\
    &\frac{\pa}{\pa \bar{t}_p} \cdot = \{ \cdot, \bh_p \}_1 = \{ \cdot , \bh_{p-1} \}_2 = \{ \cdot , \bh_{p-2} \}_3 
  \end{align}   
  for $p \geq 2$, and
  \begin{align}
    &\frac{\pa}{\pa t_1} \cdot = \{ \cdot, h_1 \}_1 = \{ \cdot , h_{0} \}_2 \\
    &\frac{\pa}{\pa \bar{t}_1} \cdot = \{ \cdot, \bh_1 \}_1 = \{ \cdot , \bh_{0} \}_2
  \end{align}   
  where the Hamiltonians are defined by \eq{qxxx} and the Poisson brackets are given in \eq{first2dexplbra}, \eq{d2pb} and \eq{ee5}.
\end{thm}

\section{Conclusion}
We have studied the two-dimensional Toda lattice hierarchy by considering an algebra of (pairs of) difference operators and, by Dirac reduction of the Poisson structures naturally associated to an $R$-matrix on such algebra, we have obtained the tri-Hamiltonian formulation of the hierarchy.

In the future we plan to study the dispersionless limit of the Poisson structures obtained here and to construct the associated infinite dimensional Frobenius manifold. In particular such dispersionless limit will be studied by applying the $R$-matrix approach considered here to a suitable algebra of Laurent series. 

It would be also interesting to consider\footnote{We are grateful to the referee for this suggestion.} the relation between the Hamiltonian formalism in the 'light-like' and the 'laboratory' coordinates.

Further generalizations to bigraded two-dimensional Toda hierarchies and their reductions are in progress.

\paragraph{\bf Acknowledgments}
The author wishes to acknowledge Prof. B. Dubrovin for many helpful and inspiring discussions.
This research was supported by EPSRC grant GR/S48424/01.

\appendix


\begin{thebibliography}{99}

\bibitem{mikhailov79}
Mikhailov, A. V. Pisma v ZhETF 30 (1979) 443.

\bibitem{ueno-takasaki84}
Ueno, Kimio; Takasaki, Kanehisa. Toda lattice hierarchy.  Group representations and systems of differential equations (Tokyo, 1982),  1--95, Adv. Stud. Pure Math., 4, North-Holland, Amsterdam, 1984.

\bibitem{mineev2000}
Mineev-Weinstein, M. ; Wiegmann, P. B.; Zabrodin, A. Phys. Rev. Lett. 84 (2000) 5106. 


\bibitem{sklyanin80}
Sklyanin, E. K. Quantum variant of the method of the inverse scattering problem. (Russian) Differential geometry, Lie groups and mechanics, III. Zap. Nauchn. Sem. Leningrad. Otdel. Mat. Inst. Steklov. (LOMI) 95 (1980), 55--128, 161;
translation in J. Soviet Math. 19 (1982), no. 5, 1546--1596.

\bibitem{semenov84}
Semenov-Tyan-Shanski\u\i, M. A. What a classical $r$-matrix is. (Russian) Funktsional. Anal. i Prilozhen.  17  (1983), no. 4, 17--33; translation in Functional Anal. Appl. 17 (1983), no. 4, 259--272.

\bibitem{oevel-ragnisco89}
Oevel, Walter; Ragnisco, Orlando. $R$-matrices and higher Poisson brackets for integrable systems. Phys. A  161  (1989), no. 1, 181--220.

\bibitem{li-parmentier89}
Li, Luen Chau; Parmentier, Serge. Nonlinear Poisson structures and $r$-matrices. Comm. Math. Phys.  125  (1989), no. 4, 545--563.

\bibitem{oevel97}
Oevel, Walter. Poisson brackets for integrable lattice systems. In: Algebraic aspects of integrable systems: in memory of Irene Dorfman. Progress in non-linear differential equations and their applications, v.26. Ed. A.S. Fokas and I.M. Gelfand, 1997.

\bibitem{kupershmidt85}
Kuperschmidt, B. A. Discrete Lax equations and differential-difference calculus. Ast\'erisque No. 123, (1985), 212 pp.
 
\bibitem{pirozerski-semenov2000}
Pirozerski, A.L. ; Semenov-Tian-Shansky, M.A. Generalized $q$-deformed Gelfand-Dickey structures on the group of $q$-pseudodifference operators. In: L.D. Faddeev Seminar on Mathematical Physics. Advances in the Mathematical Sciences, v.201, Ed. M.A. Semenov-Tian-Shansky, AMS, 2000, 321 pp. Preprint math.QA/9811025.

\bibitem{belavin-drinfeld83}
Belavin, A.A. ; Drinfel'd, V.G. Solutions of the classical Yang-Baxter equation for simple Lie algebras. (Russian) Funktsional. Anal. i Prilozhen. 16 (1982), no. 3, 1--29, 96. Translated in Functional Anal. Appl. 16 (1982) n. 3, 159--180.

\bibitem{bogoyavlensky76}
Bogoyavlensky, O.I. On perturbations of the periodic Toda lattice. Commun. Math. Phys. 51 (1976) 201--209. 

\bibitem{vanmoerbeke-mumford79}
van Moerbeke, P. ; Mumford, D. The spectrum of difference operators and algebraic curves. Acta Math. 143 (1979) 93--154.

\bibitem{adler-vanmoerbeke80}
Adler, M. ; van Moerbeke, P. Completely integrable systems, Euclidean Lie algebras, and curves. Adv. Math. 38 (1980) 267--317.

\end{thebibliography}
\end{document}